\documentclass[twocolumn,english,reprint]{revtex4}
\usepackage[T1]{fontenc}
\usepackage[latin9]{inputenc}
\usepackage{float}
\usepackage{textcomp}
\usepackage{amstext}
\usepackage{graphicx}

\makeatletter
\@ifundefined{textcolor}{}
{%
 \definecolor{BLACK}{gray}{0}
 \definecolor{WHITE}{gray}{1}
 \definecolor{RED}{rgb}{1,0,0}
 \definecolor{GREEN}{rgb}{0,1,0}
 \definecolor{BLUE}{rgb}{0,0,1}
 \definecolor{CYAN}{cmyk}{1,0,0,0}
 \definecolor{MAGENTA}{cmyk}{0,1,0,0}
 \definecolor{YELLOW}{cmyk}{0,0,1,0}
 }

\usepackage{babel}

\usepackage{babel}

\usepackage{babel}

\usepackage{babel}

\usepackage{babel}

\makeatother

\usepackage{babel}
\begin{document}

\title{Transmission electron microscopy and ferromagnetic resonance investigations of tunnel magnetic junctions using Co2MnGe Heusler alloy as magnetic electrodes}

\author{M. Belmeguenai$^{1}$, C. Genevois$^{2}$, F. Zighem$^{1}$, Y. Roussigné$^{1}$,
S-M. Chérif$^{1}$, K. Westerholt$^{3}$, A. El Bahoui$^{2}$, A.
Fnidiki$^{2}$ and P. Moch$^{1}$}

\affiliation{$^{1}$LSPM, CNRS - Université Paris XIII, 93430 Villetaneuse, France}

\affiliation{$^{2}$2Groupe de Physique des Matériaux, UMR CNRS 6634, Site Universitaire
du Madrillet, BP12, 76801 Saint Etienne du Rouvray cedex, France}

\affiliation{$^{3}$Institut für Experimentelle Physik, Ruhr-Universität Bochum,
44780 Bochum, Germany}
\begin{abstract}
High resolution transmission electronic microscopy, nano-beam electronic
diffraction, energy dispersive X-rays scanning spectroscopy, Vibrating
Sample Magnetometry (VSM) and FerroMagnetic Resonance (FMR) techniques
are used in view of comparing (static and dynamic) magnetic and structural
properties of Co$_{2}$MnGe (13 nm)/Al$_{2}$O$_{3}$ (3 nm)/Co (13
nm) tunnel magnetic junctions (TMJ), deposited on various single crystalline
substrates (a-plane sapphire, MgO(100) and Si(111)). They allow for
providing a correlation between these magnetic properties and the
fine structure investigated at atomic scale. The Al$_{2}$O$_{3}$
tunnel barrier is always amorphous and contains a large concentration
of Co atoms, which, however, is significantly reduced when using a
sapphire substrate. The Co layer is polycrystalline and shows larger
grains for films grown on a sapphire substrate. The VSM investigation
reveals in-plane anisotropy only for samples grown on a sapphire substrate.
The FMR spectra of the TMJs are compared to the obtained ones with
a single Co and Co$_{2}$MnGe films of identical thickness deposited
on a sapphire substrate. As expected, two distinct modes are detected
in the TMJs while only one mode is observed in each single film. For
the TMJ grown on a sapphire substrate the FMR behavior does not significantly
differ from the superposition of the individual spectra of the single
films, allowing for concluding that the exchange coupling between
the two magnetic layers is too small to give rise to observable shifts.
For TMJs grown on a Si or on a MgO substrate the resonance spectra
reveal one mode which is nearly identical to the obtained one in the
single Co film, while the other observed resonance shows a considerably
smaller intensity and cannot be described using the magnetic parameters
appropriate to the single Co$_{2}$MnGe film. The large Co concentration
in the Al$_{2}$O$_{3}$ interlayer prevents for a simple interpretation
of the observed spectra when using Si or MgO substrates.
\end{abstract}


\maketitle

\section{Introduction}

Spin electronics is an emerging technology exploiting the spin of
electron as an information carrier. Tunnel magnetic junctions (TMJ),
consisting of a tunnelling barrier sandwiched between two ferromagnetic
electrodes, belong to devices presenting a great interest due to their
use in magnetic memories {[}1{]}, in low field magnetic sensors {[}2{]}
and in microwave components for spintronic applications {[}3{]} for
which high tunnel magnetoresistance ratios (TMR) are highly desirable.
TMR is directly related to the spin polarization of ferromagnetic
electrodes {[}4{]}. Therefore, half metallic materials such as Heusler
alloys, such as NiMnSb, PtMnSb and Co-based Heusler compounds like
Co$_{2}$MnSi and Co$_{2}$MnGe, should be ideal compounds as high
spin polarized current sources allowing for realizing large TMR values.
The efficiency of a TMJ, and therefore its TMR value, strongly depend
on interfacial roughness, on inter-diffusion and on oxygen content,
which in turn depend on the materials used in the stack and on the
conditions of deposition and of annealing. Furthermore, the substrate
material as well as its orientation have an impact on the magnetic
anisotropy of the magnetic thin films, and thus on the TMR value,
because of the band hybridization and of the spin-orbit interaction
at the interface. Therefore, the control of such parameters should
allow for enhancing the half-metallicity of the electrodes and thus
for increasing the TMR values. Co-based Heusler alloys {[}5, 6{]}
are promising materials for spintronic applications, because a number
of them possess a high Curie temperature {[}6{]} and, therefore, may
consist in alternative compounds to obtain half metallicity even at
room temperature. These materials have been used as electrodes {[}7-10{]}
in TMJs where TMR up to 360$\%$ has been demonstrated at room temperature
{[}9, 10{]}. However, up to now, the demonstrated TMR amplitudes using
Heusler alloys as magnetic electrodes in TMJs remain lower than the
extremely large TMR at room temperature which has been predicted and
demonstrated in TMJs using (001)-textured MgO as a tunnel barrier
and normal transition metals or their alloys as electrodes {[}11,
12{]}. Apart from high TMR values, it is also important to understand
the magnetization dynamics and the magnetic anisotropy of TMJs in
relation with the interfacial characteristics for realizing high speed
spintronic devices. Therefore, ferromagnetic resonance (FMR), vibrating
sample magnetometer (VSM), ultrahigh resolution and scanning transmission
electronic microscopy (HRTEM and STEM), nano-beam electronic diffraction
and energy dispersive X-rays scanning spectroscopy (STEM-EDX) have
been used for the investigation of static and dynamic magnetic properties
in correlation with the atomic scale characterization of interfaces
involved in Co$_{2}$MnGe (13 nm)/Al$_{2}$O$_{3}$ (3 nm)/Co (13
nm) TMJs deposited on a-plane sapphire, on MgO and on Si substrates.
The properties of a Co$_{2}$MnGe/Al$_{2}$O$_{3}$/Co TMR device
have been previously discussed by Verduijn et al. {[}13{]} and compared
to those of a Co$_{2}$MnGe/Al$_{2}$O$_{3}$/Co TMR element prepared
using identical process parameters. They found a 27$\%$ TMR value
at 77 K for Co$_{2}$MnGe/Al$_{2}$O$_{3}$/Co, slightly larger than
for a Co$_{2}$MnGe/Al$_{2}$O$_{3}$/Co junction. However, at room
temperature, the TMR value of Co$_{2}$MnGe/Al$_{2}$O$_{3}$/Co drastically
decreases to 9\%. Therefore, in order to understand the reasons of
this low TMR value at room temperature, magnetic properties are studied
in this paper in close relation with the microstructural properties
of Co$_{2}$MnGe/Al$_{2}$O$_{3}$/Co TMJ.

\section{Samples preparation and experimental methods}

The Co$_{2}$MnGe(13nm)/Al$_{2}$O$_{3}$(3nm)/Co(13nm) TMJs were
prepared in-situ under ultrahigh vacuum (UHV) conditions (base pressure
$10{}^{-8}$ Pa) using a combination of ion beam sputter deposition
(IBSD) and UHV-magnetron sputtering on sapphire a-plane, Si(111) and
MgO(100) substrates. The Co$_{2}$MnGe layer was grown at a substrate
temperature of $300\text{\textdegree}$C on a 4 nm thick Vanadium
seed layer by UHV magnetron sputtering at a rate of 0.015 nm/s. After
the deposition of the Heusler layer, the substrate was cooled down
to room temperature, transferred to the IBSD chamber and then cleaned
by ion beam etching with a 150 eV Ar-ion beam to remove any surface
layer, which might have been oxidized during cooling down of the substrate.
The Al$_{2}$O$_{3}$-layer is created by depositing 1.5 nm Al at
room temperature at a sputtering rate of 0.08 nm/s and a subsequent
RF plasma oxidation process in a separate oxidation chamber. The oxidation
chamber was filled with pure oxygen gas at a pressure of 6 kPa, and
the Al-layer was oxidized due to the applied RF-power of 5 W during
30 s. In a final step the Co-layer was deposited by IBSD with the
substrate at room temperature at a sputtering rate of 0.03 nm/s and
capped by a 4 nm Au protective layer. More details about the fabrication
procedure can be found in {[}13{]}.

The Co and Co$_{2}$MnGe single layers, which serve as reference samples
below, were prepared using the same procedures and keeping all parameters
identical to those used during the preparation of the TMJs.

The characterization of the microstructures and elemental compositions
of the TMJ grown on a-plane sapphire, MgO(100) and Si(111) substrates,
were performed using a transmission electron microscope. Three TEM
cross sections are prepared by in situ lift out using a Zeiss scanning
electron microscope equipped with a Focused Ion Beam. To avoid damage
from the high-energy ion beam during sample preparation, a platinum
(Pt) layer was deposited to protect the sample surface. High resolution
transmission electron microscopy and scanning transmission electron
microscopy observations were carried out on a JEOL JEM ARM 200F (JEOL
Ltd.) operating at 200 kV. This microscope was equipped with a field
emission gun and an aberration (Cs) corrector on the electron probe.
High angle annular dark field STEM (HAADF-STEM) images were acquired
with a camera length of 8 cm and a probe size of 0.1 nm. The contrast
of these micrographs is linked to the atomic number Z of the phase
and thus permits to obtain a chemical contrast. Elemental compositions,
with a precision of more or less 2 atom per cent (at.$\%$), were
performed by STEM-EDX using a JEOL detector with a probe size of 0.4
nm.

Their magnetic dynamic properties have been studied by micro-strip
ferromagnetic resonance (MS-FMR). The MS-FMR characterization was
done with the help of a field modulated FMR setup using a vector network
analyser (VNA) operating in the 0.1-40 GHz frequency range. The sample
(with the film side in direct contact) is mounted on a 0.5 mm micro-strip
line connected to the VNA and to a lock-in amplifier to derive the
field-modulated measurements via a Schottky detector. This setup is
piloted via a Labview program providing flexibility of a real time
control of the magnetic field sweep direction, step and rate, real
time data acquisition and visualization. It allows both frequency
and field-sweeps measurements with magnetic fields up to 2 T applied
parallel or perpendicular to the sample plane. In-plane angular dependence
of resonance frequencies and fields are used to measure anisotropies.
The complete analysis of in-plane and perpendicular field resonance
spectra exhibiting the excited modes leads to the determination of
most of the magnetic parameters: effective magnetization, gyromagnetic
factor, interlayer exchange coupling and anisotropy terms.

\section{HRTEM and STEM characterization}

Figure1 shows the TMJ structure for the three samples. These structures
consist in a stack of layers with different elemental compositions.
On the STEM-HAADF micrograph, larger is the average atomic number
of the layer, brighter it appears. The three samples present a similar
structure with the succession of the following layers: substrate/V/Co$_{2}$MnGe/Al$_{2}$O$_{3}$/Co/Au.
The interfaces seem to be well defined and flat, except for the Co/Au
one, which is very rough and wavy. The Au layer is not uniform and
presents some holes. The average thicknesses of the V, Al$_{2}$O$_{3}$
and Co layers do not depend on the different substrates and are respectively:
4 nm, 3-4 nm and 8-11 nm. An example of thickness measurement is presented
on figure 1d. The variation of the Co layer thickness reflects the
roughness of its interface with the gold layer. Concerning the thickness
of the Co$_{2}$MnGe layer, it depends on the substrate: it is of
about 13 nm for the sapphire and Si substrates but of only 10 nm for
the MgO substrate. The thickness measurements have been carried out
on the STEM micrographs with a precision of 1 nm.

The dependence of the elemental composition versus the coordinate
normal to the sample was obtained for the three TMJs using EDX line
scans. The results are shown on figure 2. In each layer, mainly due
to inter-diffusion, the elemental concentration differs from the expected
nominal one and is not uniform. The detailed mapping of the elemental
composition depends on the substrate. Notice that the thin film of
Al$_{2}$O$_{3}$ separating the two magnetic layers contains a very
high concentration of Co in the case of MgO or of Si substrates (average
value overpassing 40 at.$\%$) and a rather smaller one (average value
overpassing 20 at.$\%$) for a sapphire substrate. The effective structure
is poorly approximated by an abrupt Co$_{2}$MnGe/Al$_{2}$O$_{3}$/Co
trilayers arrangement, except, maybe, in the case of a sapphire substrate.
In addition, with the MgO substrate, the Mn concentration in the Heusler
film lies well below the expected value (7 at.$\%$ instead of 25
at.$\%$). More generally, accurate EDX line scans (see Figures 3
and 4) show that around every interface there is an important inter-diffusion
of the constitutive elements of adjacent layers: as a result the elemental
composition of each layer is strongly modified near its interfaces
on a distance of about 2 nm. Due to the small thickness (3 nm) of
the Al$_{2}$O$_{3}$ film some alteration of the properties of the
studied TMJs is expected.

HRTEM micrographs and nano-beam diffractions have permitted to check
the crystalline quality of the different layers and the epitaxial
relationships. The substrate and the vanadium layer are monocrystalline
and the vanadium layer shows a perfect epitaxy with the substrate.
This is illustrated, in the case of the MgO substrate, by a micrograph
in figure 5a and by two diffraction patterns in Figures 6a and 6b
related to MgO and to the vanadium layer, respectively: the ($110$)V
plane of the vanadium layer coincides with the ($100$)MgO plane of
the substrate, as expected. The diffraction patterns shown in Figure
6c and in Figure 6d concern the Co$_{2}$MnGe layer, near of the V/Co$_{2}$MnGe
interface and in the vicinity of its centre, respectively: the rather
good epitaxy observed near the interface does not persist in the central
region which shows a polycrystalline distribution, as attested by
the presence of different diffraction spots arising from many different
nano-grains. For the three studied TMJs, the Al$_{2}$O$_{3}$ barrier
is amorphous and the Co layer is found to be polycrystalline, as resulting
from multiple diffraction spots arising from its central region (Figure
6e). However, for the TMJ grown on a sapphire substrate, the mean
size of the nano-grains is substantially increased, allowing for observing
nano-beams diffraction patterns related to only one grain, as shown
in Figure 6f.

In conclusion, the studied TMJs show an important inter-diffusion
between their constitutive layers, and present polycrystalline magnetic
films. However, the use of a sapphire substrate seems to provide the
best choice in view of obtaining TMJs of good quality (smallest diffusion
and largest nano-grains).

\section{Magnetic properties}

The experimental magnetic dynamic data have been analysed considering
a magnetic energy density characterized by Zeeman, demagnetizing and
anisotropy contributions given by equations (1) and (2) respectively
for Co2MnGe and Co respectively:

$E_{A}=-\mu_{0}M_{A}H(\sin\theta_{MA}\sin\theta_{H}\cos(\varphi_{MA}-\varphi_{H})+\cos\theta_{MA}\cos\theta_{H})-\left(\frac{\mu_{0}}{2}M_{A}^{2}-K_{\bot A}\right)\sin^{2}\theta_{MA}-\frac{1}{2}(1+cos(2(\varphi_{MA}-\varphi_{uA}))K_{uA}\sin^{2}\theta_{MA}-\frac{1}{8}(3+\cos4(\varphi_{MA}-\varphi_{4}))K_{4}\sin^{4}\theta_{MA}$(1)

$E_{B}=-\mu_{0}M_{B}H(\sin\theta_{MB}\sin\theta_{H}\cos(\varphi_{MB}-\varphi_{H})+\cos\theta_{MB}\cos\theta_{H})-\left(\frac{\mu_{0}}{2}M_{B}^{2}-K_{\bot B}\right)\sin^{2}\theta_{MB}-\frac{1}{2}(1+cos(2(\varphi_{MB}-\varphi_{uB}))K_{uB}\sin^{2}\theta_{MB}$
(2)

In the above expressions, $H$ is the applied magnetic field; the
other parameters stand for the saturation magnetization ($M$), for
the perpendicular anisotropy contribution ($K_{\perp}$), for the
two-fold ($Ku$, $\varphi_{u}$) and the four-fold in-plane anisotropy
contributions ($K_{4}$$\varphi_{4}$), using the additional suffixes
labeling the concerned layer ($A$ for Co$_{2}$MnGe or $B$ for Co).
The out-of-plane ($\theta$) and in-plane directions ($\varphi$)
also appear with the appropriate indices. Tentatively, the total magnetic
energy of the TMJ consisting in two magnetic layers of thicknesses
$d_{A}$ and $d_{B}$ can be written as:

$E_{B}=d_{A}E_{A}+d_{B}E_{B}-J_{1}\left(\sin\theta_{MA}\sin\theta_{MB}\cos(\varphi_{MA}-\varphi_{MB})+\cos\theta_{MA}\cos\theta_{MB}\right)$
(3)

Above, $J_{1}$ defines a bilinear exchange coupling between the two
films.

In an applied magnetic field $H$ two uniform magnetic modes result
from the above expression of the energy: their frequencies are obtained
in the usual way by solving the equations of motion of the magnetizations
{[}14{]} around their equilibrium positions (these positions are obtained
by minimizing $E$). Indeed, in the absence of coupling ($J_{1}=0$),
the two frequencies do not differ from their values in each single
layer (which are proportional to the gyromagnetic factor $\gamma_{A}$
and $\gamma_{B}$, for $A$ and $B$, respectively). Conversely, at
a fixed frequency one expects for two different resonant in-plane
magnetic fields applied in any defined direction. In the following,
as usual, we introduce, with the appropriate suffixes $A$ or $B$,
the effective Landé factors $g\left(\frac{\gamma}{2\pi}\right)=g\frac{e}{4\pi m_{e}}=g\times13.996$
GHz/T, where $e$ is the absolute value of the electron charge and
$m_{e}$ is its mass), the effective demagnetizing fields ($\mu_{0}M_{eff}=\mu_{0}M-\frac{2K_{\perp}}{M}$),
the uniaxial in-plane anisotropy field $\mu_{0}H_{u}=\frac{2K_{u}}{M}$and
the fourfold in-plane anisotropy field $\mu_{0}K_{4}=\frac{2K_{4}}{M}$.

\subsection{Static magnetic measurements}

The coercive fields are strongly dependent function of the anisotropy
fields, of the magnetostatic interaction and of the exchange interaction
between the magnetic films. All these quantities depend on the chemical
composition, on the morphology and on the crystalline structure of
the thin films. For all the studied samples the hysteresis loops were
obtained by VSM with an in-plane magnetic field applied along various
orientations $\varphi_{H}$ (where $\varphi_{H}$ is the in-plane
angle between the magnetic applied field $H$ and one of the edges
of the substrate). Figure 7 shows representative behaviors of these
samples. The observed shape mainly depends on the field orientation
both for single layers and TMJ grown on sapphire, in agreement with
the presence of in-plane anisotropy. In contrast, the hysteresis loops
of TMJs grown on Si and MgO do not depend on $\varphi_{H}$, suggesting
the absence of anisotropy (Fig. 7e and 7f). The magnetization loops
along the easy axis are shown on figure 7a for the two single layers
and for the three TMJs. The $\mu_{0}H_{C}$ values lie near 2.3 mT,
except for the Co single layer, which presents a higher $\mu_{0}H_{C}$
(4.8 mT). Figure 7b illustrates the high in-plane anisotropy observed
in the Co single layer: along the hard axis $\varphi_{H}=90\text{\textdegree}$
the saturation field for magnetization is 30 mT). The presence of
Co magnetic anisotropy in TMJ on sapphire substrate and its disappearance
in TMJs on Si and MgO substrates suggest that this uniaxial anisotropy
is most probably induced by the interface with the sapphire substrate
{[}15{]}. As confirmed below by FMR, the hysteresis loops of the Co$_{2}$MnGe
single layer (Fig. 7c) reveal that its planar anisotropy consists
of the superposition of a fourfold contribution, with easy axes parallel
to the substrate edges, and of a uniaxial term. The fourfold anisotropy
is presumably of magnetocrystalline nature while we attribute the
uniaxial contribution at least partially originating from a slight
misorientation of the surface of the substrate as discussed in details
in {[}16, 17{]}. Finally, the hysteresis loop of the TMJ grown on
sapphire presents a narrow plateau at small applied fields, suggesting
that Co and Co$_{2}$MnGe layers (Fig. 7d) are uncoupled, and that
their coercive fields are very close from each other, which implies
a significant decrease of the Co anisotropy, compared to the single
Co layer. However, for TMJs on Si and on MgO, the absence of anisotropy
for the two magnetic layers explains the complete disappearance of
any double switching behaviour. The presence of anisotropy in TMJ
grown on sapphire is presumably related to their best quality, as
shown by the above microstructural characterization.

In addition, variations of the reduced remanent magnetization ($M_{r}/M$),
as function of $\varphi_{H}$, of the single layers are also shown
in view of comparison with other results discussed in the next section.

\subsection{Dynamic measurements}

In a first step the magnetic parameters of the individual layers were
derived with the help of the FMR study of our single magnetic Co and
Co$_{2}$MnGe films. In these films, only one resonance mode is observed,
as expected. Figures 8a and 8b illustrate the experimental in-plane
angular dependence of the frequency of the uniform mode for Co and
Co$_{2}$MnGe layers respectively, compared to our best fits using
the model above described. The derived values of the magnetic parameters
corresponding to these best fits are reported in Table I. Notice that
a uniaxial term well describes the in-plane anisotropy in the Co sample
while both uniaxial and fourfold in-plane anisotropy terms are requested
to give account for the data concerning Co$_{2}$MnGe. In addition,
the derived in-plane anisotropy characteristics are consistent with
the above mentioned angular variation of the reduced remanent magnetization
($M_{r}/M$).

In the TMJs, two resonance modes should be present in the FMR spectra.
For sweep frequency measurements using an in-plane applied magnetic
field, such a twofold behavior is only observed for the TMJ grown
on sapphire (Fig. 9a) while the TMJs grown on MgO or on Si show a
unique resonance line. The indices 1 and 2 are used to refer to the
lower and to the higher frequency mode present with a sapphire substrate,
respectively. In the case of a MgO or of a Si substrate the resonance
frequency lies near of the value observed for the mode 2 in the TMJ
deposited on sapphire. Sweep field measurements (Fig. 9b) allow for
the detection of the two modes whatever is the substrate, maybe due
to the best sensitivity. However, compared to the case of sapphire,
the intensity of mode 1 is significantly reduced with MgO or Si substrates.
Let us remember that it results from the STEM investigation that with
Si or MgO, the Al$_{2}$O$_{3}$ interlayer contains a high concentration
of inter-diffused Co: the description of the studied structure in
terms of a junction between two well separated homogeneous magnetic
layers then becomes rather poor. On the other hand, it is probable
that the best quality of Co$_{2}$MnGe layers is obtained using a
sapphire substrate.

The field dependences of the frequencies of the observed modes for
the three TMJs, compared to those of the single layers, are shown
on Figure 10. Assuming that the exchange coupling does not induce
very strong perturbations among the two modes, one of them represents
a Co-like uniform magnetic excitation and the other one represents
a Heusler-like uniform magnetic excitation. Due to the field dependence
of mode 2 it is assigned to consist in the Co-like uniform mode: a
preliminary derivation neglecting any coupling leads an effective
$g$-factor of 2.17 and to a demagnetizing field ranging in the (1.47-1.54)
T interval, to compare the 2.17 and 1.55 T values respectively found
for the single Co layer. Indeed, mode 1 is identified as the Heusler-like
one: for the TMJ grown on a sapphire substrate, the same approximation
leads to $g=2.02$ and to a demagnetizing field of 0.95 T, identical
to the values found for the single Heusler layer. However, in the
case of MgO or of Si substrates the demagnetizing field is substantially
reduced: the highly Co concentrated interlayer seems to induce a perpendicular
anisotropy but the interpretation of the result is missing. Concerning
the in-plane anisotropy, it was derived for mode 2 in the three available
TMJs through the in-plane angular dependence of the FMR spectra: in
the case of an MgO or of a Si substrate it vanishes; with a sapphire
substrate (Figure 8c) it consists into a rather small uniaxial contribution
($\mu_{0}H_{u}$ = 1.1 mT). Notice that the single Co layer is affected
by a high uniaxial contribution (30 mT, see Figure 8a): this put in
evidence the difference between the (crystalline sapphire)-Co and
the (amorphous alumina)-Co interfacial interaction. The in-plane anisotropy
relative to mode 1 could be only studied in the TMJ grown on a sapphire
substrate (Fig. 8b): it can be described as deriving from the superposition
of a four-fold symmetry term and of a uniaxial one, a behaviour analogous
to the observed one in the single Heusler layer (Figure 8b).

Finally, it appears that the fit of our data cannot be improved by
taking in account the exchange interaction: a large exchange would
induce similar shifts for mode 1 and mode 2, in contrast with the
observed results. In view of the other variations of the parameters
monitoring the resonance the exchange is too small to be derived from
the FMR studies. However, this small exchange can be detected from
the analysis of the static magnetic properties.

\section{Conclusion}

Co$_{2}$MnGe(13nm)/Al$_{2}$O$_{3}$(3 nm)/Co(13nm) TMJs deposited
on a-plane sapphire, MgO and Si substrates have been prepared. For
comparison, single films of Co$_{2}$MnGe and of Co with a thickness
of 13 nm have been also deposited on sapphire. Their nanoscale structure,
compared to their static and dynamic magnetic properties, have been
studied. The ultrahigh resolution transmission electron microscopy
revealed differences between the TMJs which have a direct impact on
their magnetic characteristics. The FMR measurements show the presence
of two modes deriving from the Co and from the Co$_{2}$MnGe layer
which are only very weakly coupled. The magnetic parameters of these
modes depend on the quality of the TMJ and more or less differ from
those of the single layers.

\begin{figure}
\includegraphics[bb=40bp 111bp 220bp 590bp,clip,width=8.5cm]{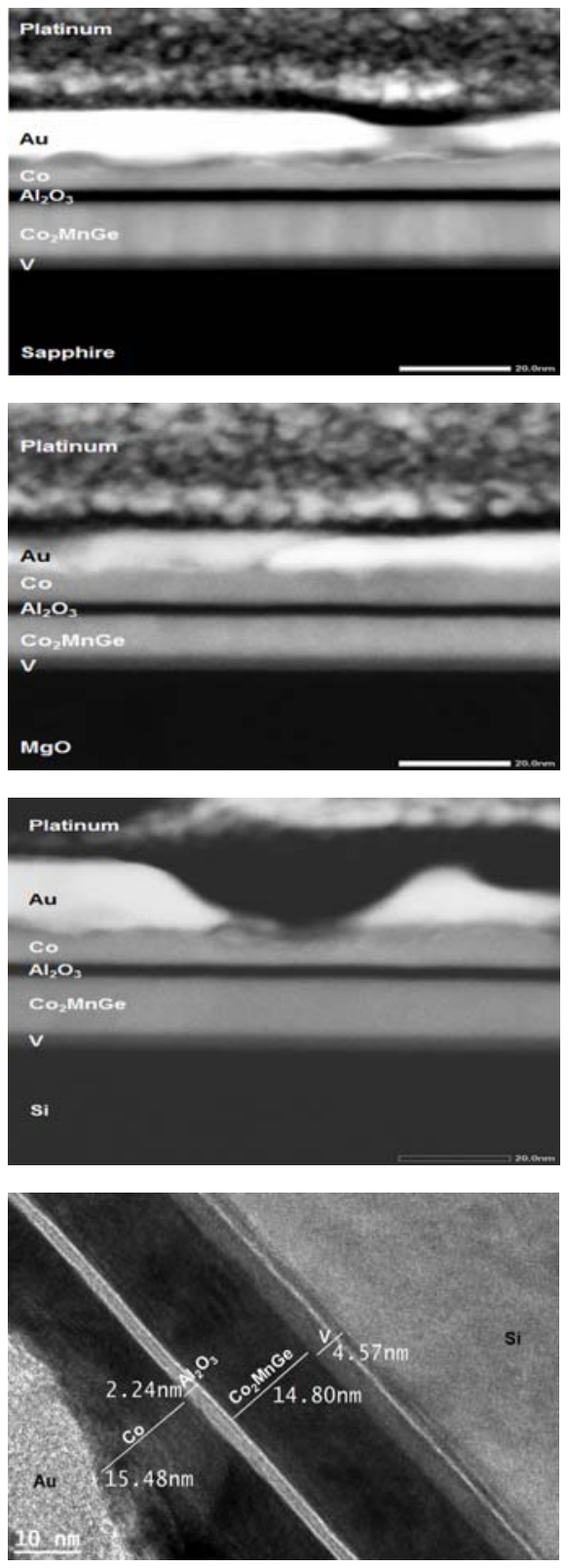}

\caption{Cross sectional STEM-HAADF micrographs of the TMJ grown on (a) a-plane
sapphire, (b) MgO(100) and (c) Si(111) substrates. (d) Example of
thickness measurements for TMJ grown on Si. }
\end{figure}

\begin{figure}
\includegraphics[bb=20bp 250bp 280bp 595bp,clip,width=8.5cm]{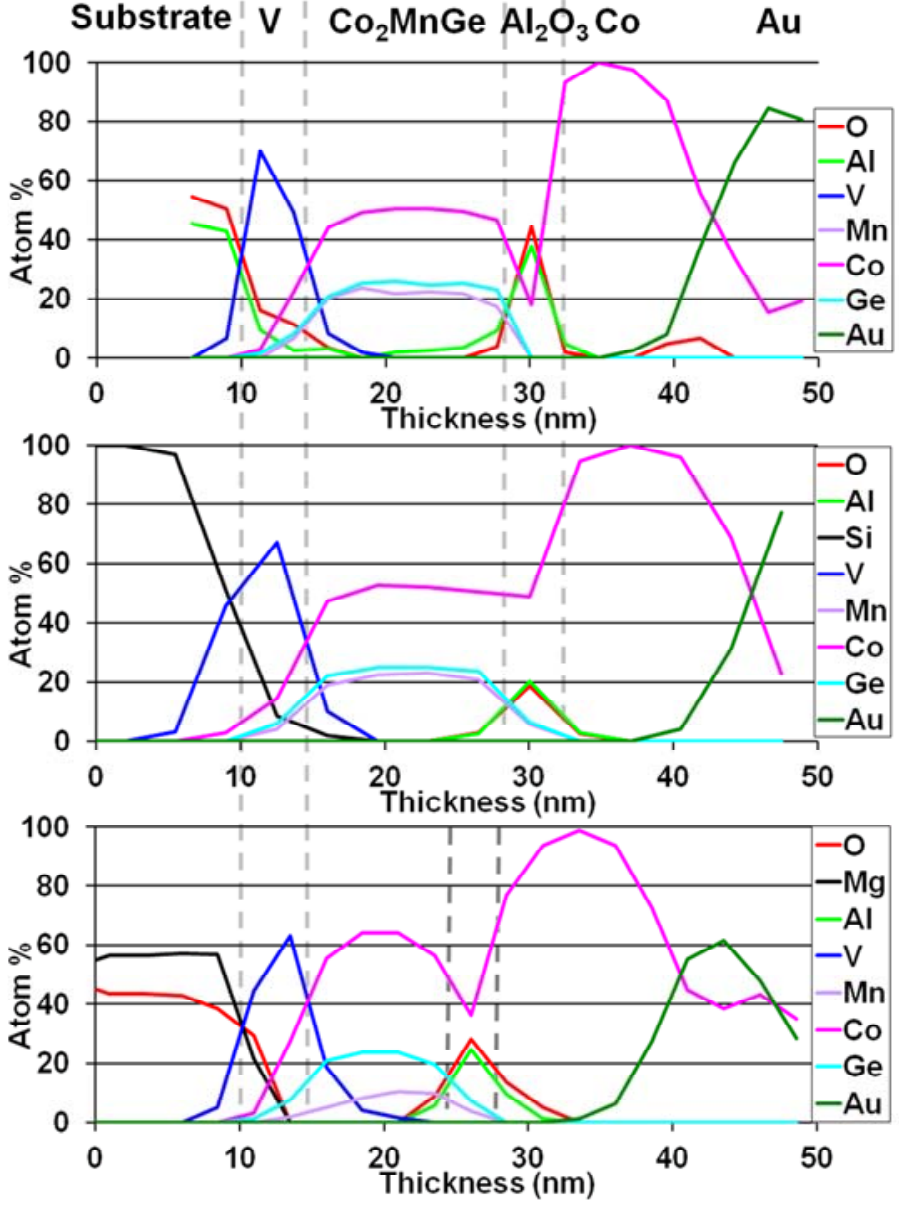}

\caption{(Colour online) Evolution of the elemental composition through the
TMJ grown on a-plane sapphire, Si(111) and MgO(100) substrates. These
results are obtained thanks to EDX line scans.}
\end{figure}

\begin{figure}
\includegraphics[bb=20bp 250bp 280bp 595bp,clip,width=8.5cm]{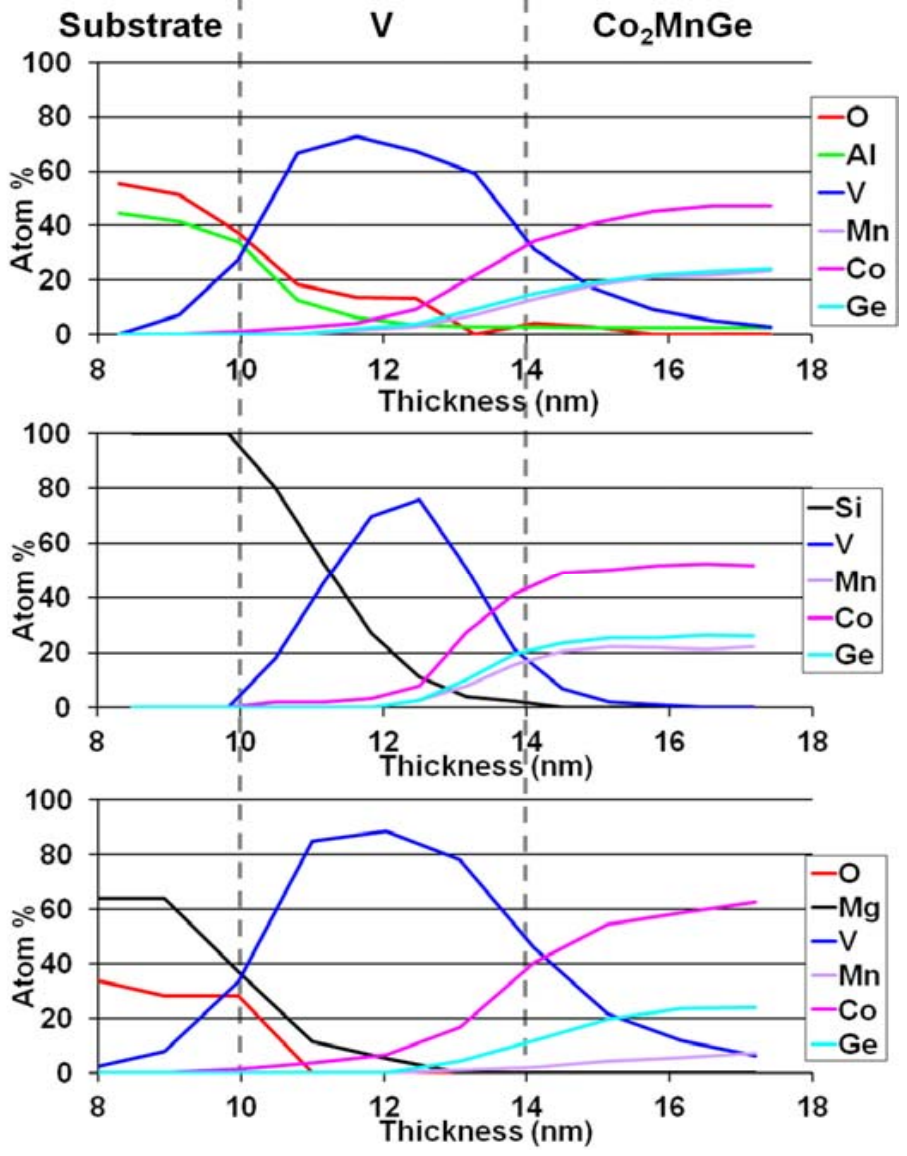}

\caption{(Colour online) Evolution of the elemental composition through the
vanadium/Co$_{2}$MnGe interface on the TMJ grown on a-plane sapphire,
Si(111) and MgO(100) substrates. These results are obtained thanks
to EDX line scans.}
\end{figure}

\begin{figure}
\includegraphics[bb=20bp 250bp 280bp 595bp,clip,width=8.5cm]{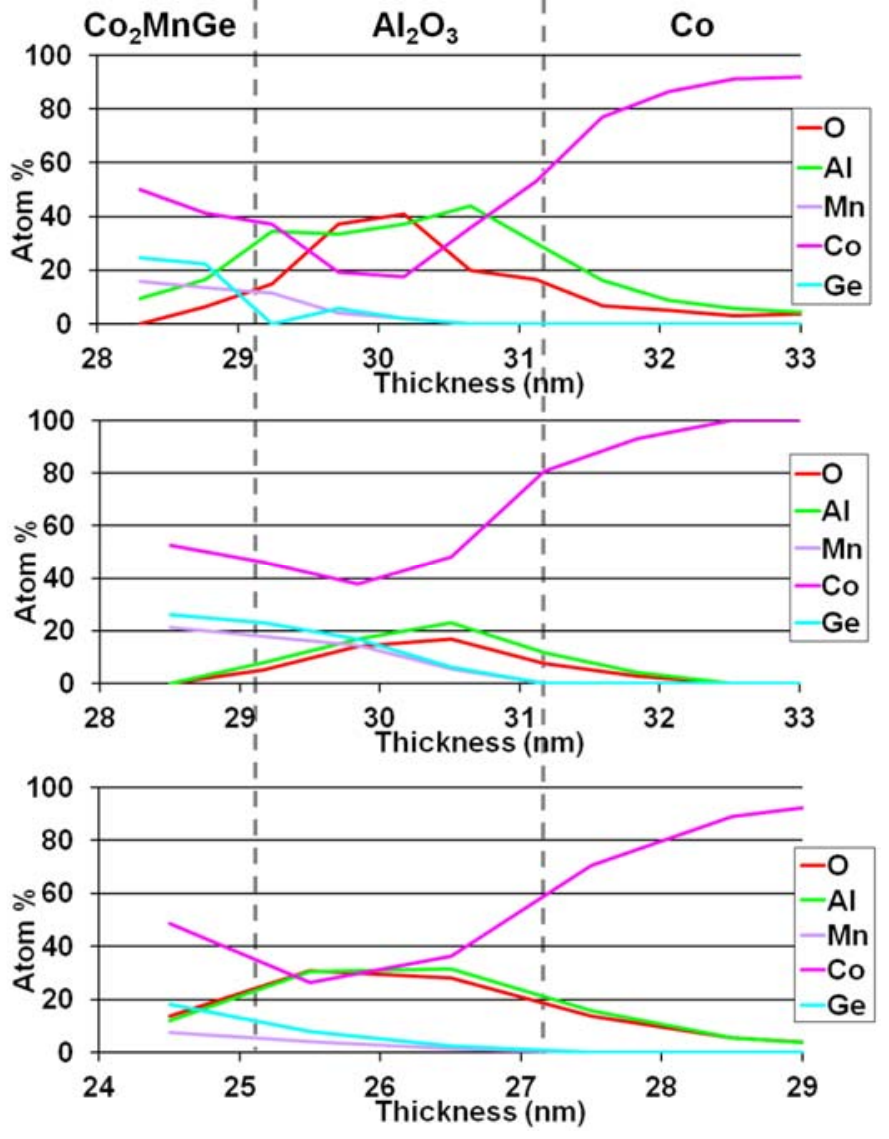}

\caption{(Colour online) Evolution of the elemental composition through the
Co$_{2}$MnGe/Al2O3/Co interfaces on the TMJ grown on a-plane sapphire,
Si(111) and MgO(100) substrates. These results are obtained thanks
to EDX line scans.}
\end{figure}

\begin{figure}
\includegraphics[bb=30bp 200bp 280bp 590bp,clip,width=8.5cm]{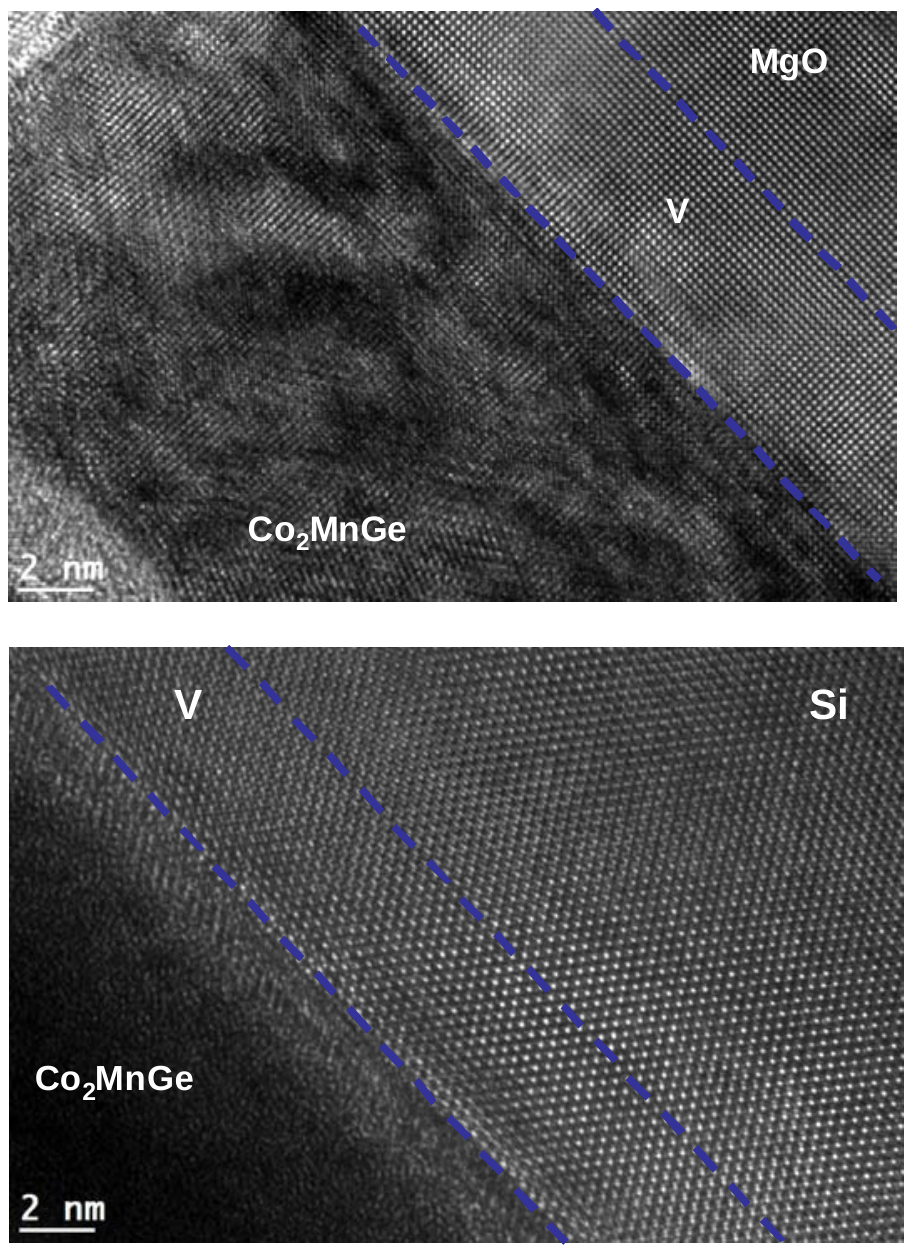}

\caption{(Colour online) Cross sectional HRTEM micrograph of the TMJ grown
on (a) MgO(100) and (b) Si(111) substrates showing the epitaxial growth
between the vanadium layer and the substrate and between the vanadium
layer and the Co$_{2}$MnGe layer, respectively. Figure 6: (a) to
(e): nano-beam diffractions in the TMJ grown on a MgO(100) substrate}
\end{figure}

\begin{figure}
\includegraphics[bb=30bp 150bp 380bp 590bp,clip,width=8.5cm]{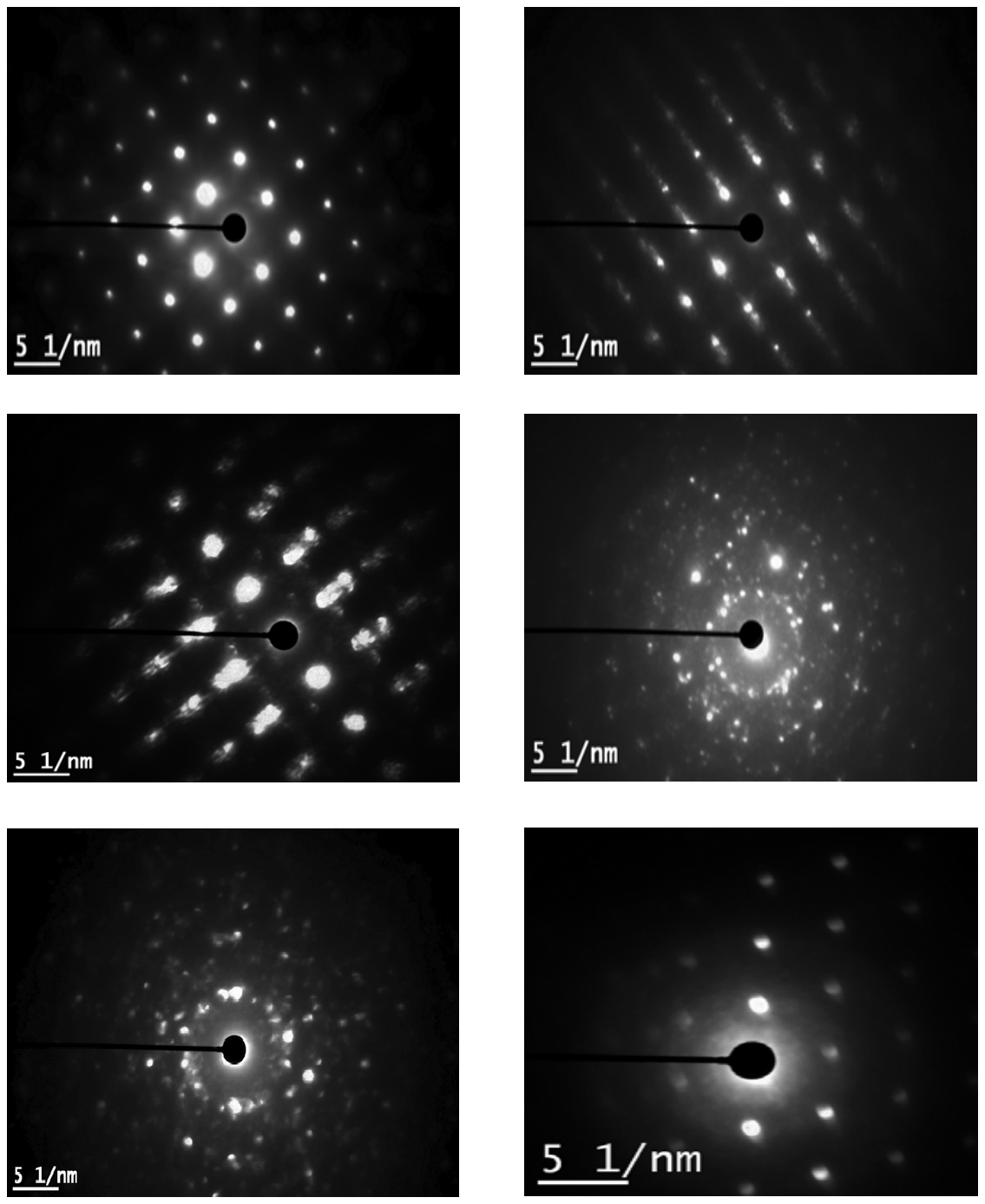}

\caption{(a) to (e): nano-beam diffractions in the TMJ grown on a MgO(100)
substrate: diffraction patterns arising from (a) the substrate layer,
(b) the vanadium layer, (c) the Co$_{2}$MnGe layer located close
to the vanadium side, (d) the center of the Co$_{2}$MnGe layer, (e)
the center of the Co layer. (f): nano-beam diffraction pattern acquired
at the center of the Co layer of the TMJ grown on the a-plane sapphire.}
\end{figure}

\begin{figure}
\includegraphics[bb=30bp 0bp 550bp 595bp,clip,width=8.5cm]{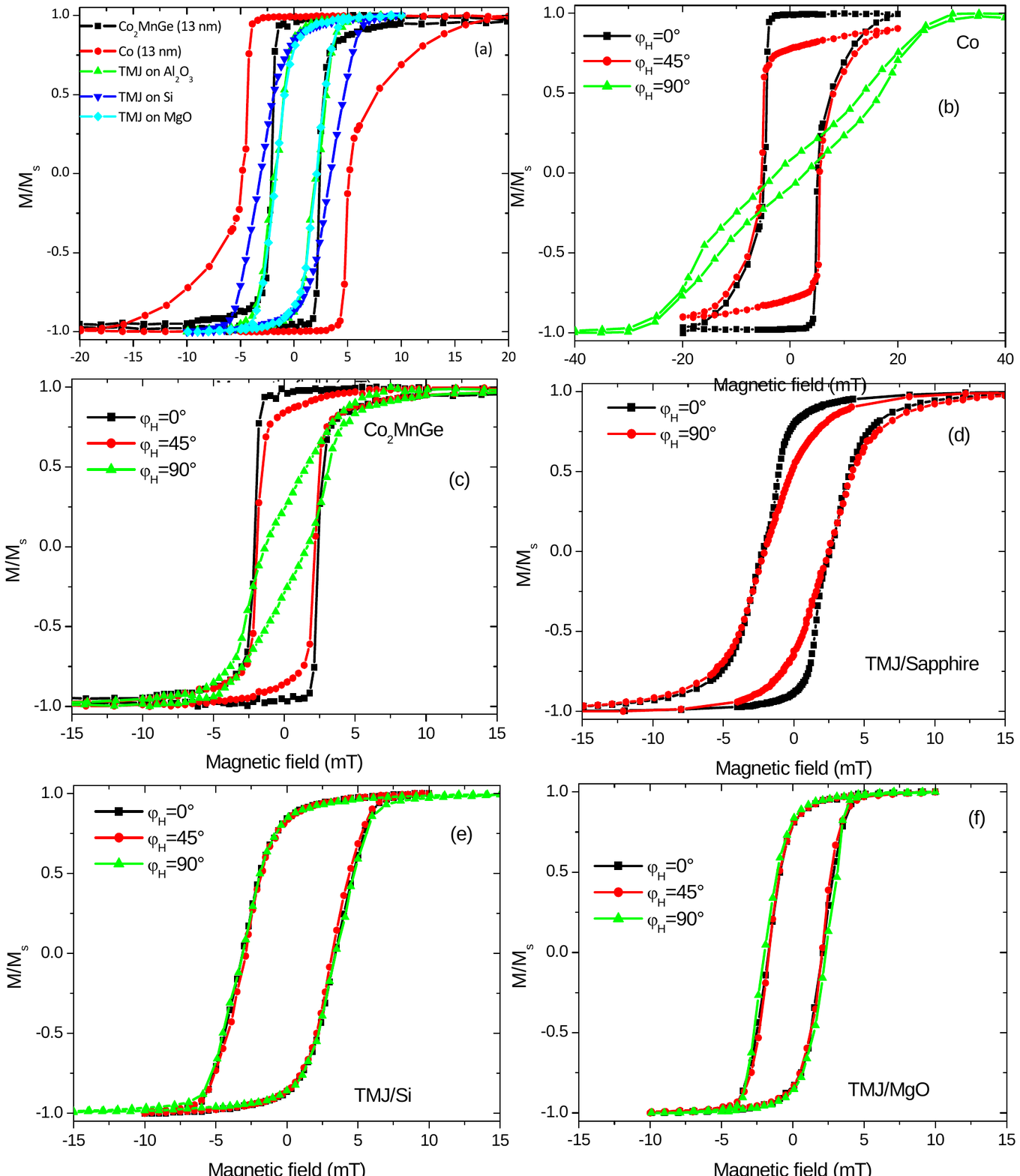}

\caption{(Colour online) (a) Easy axis VSM magnetization loops of the different
studied samples and VSM hysteresis loops for an applied magnetic field
at various angles ($\varphi_{H}$) with the substrate edges (b) Co
single layer, (c) Co$_{2}$MnGe single layer, (d) TMJ grown on Sapphire,
(e) TMJ grown on Si and (f) TMJ grown on MgO.}
\end{figure}

\begin{figure}[H]
\includegraphics[bb=20bp 0bp 300bp 595bp,clip,width=8.5cm]{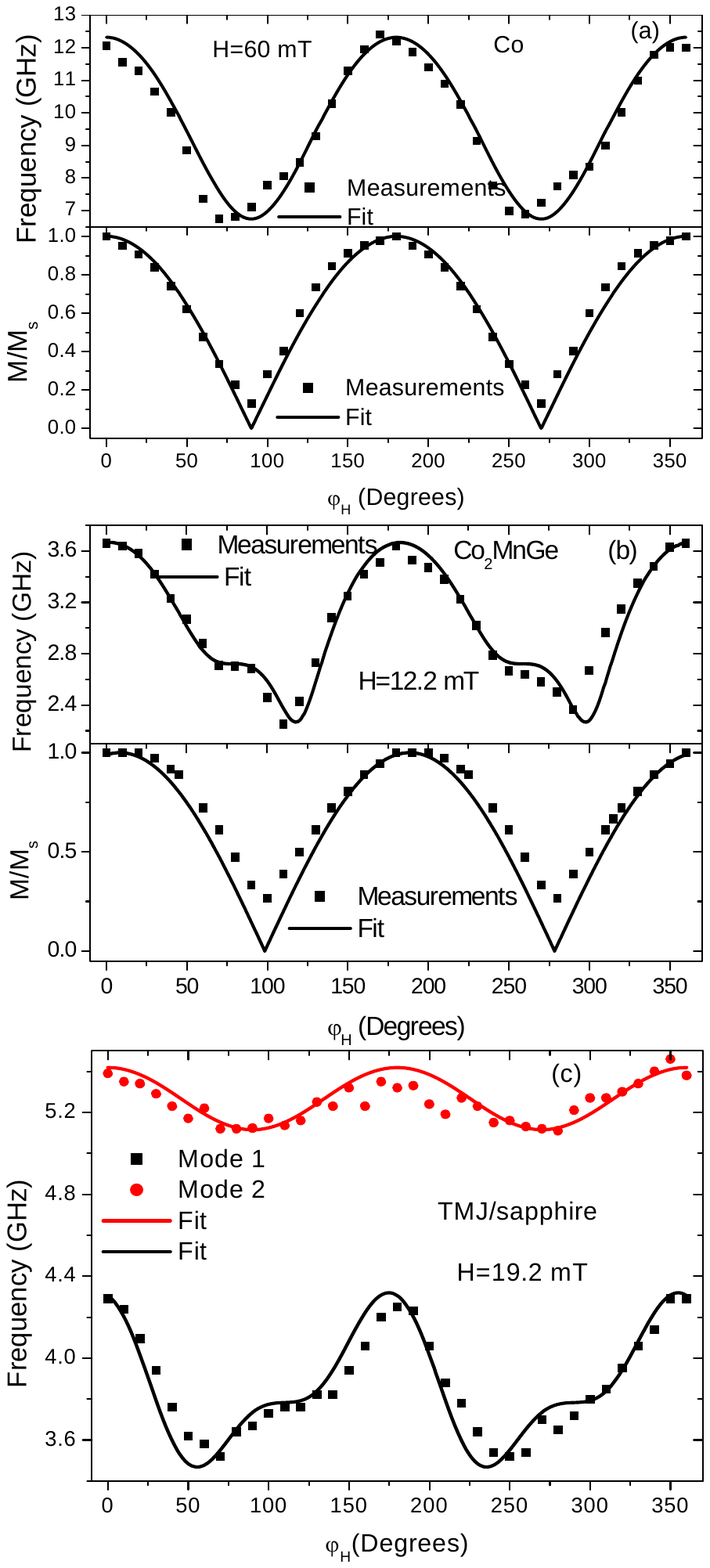}

\caption{(Colour online) In-plane angular dependences of the resonance frequency
of modes 1 and 2 and reduced remanent magnetization of the (a) Co,
(b) Co$_{2}$MnGe single layers and (c) Co$_{2}$MnGe(13nm)/Al$_{2}$O$_{3}$(3nm)/Co(13
nm) TMJ grown on a-plane sapphire substrate. The full lines are obtained
using the model presented in the text with the parameters indicated
in Table I.}
\end{figure}

\begin{figure}
\includegraphics[bb=20bp 160bp 300bp 595bp,clip,width=8.5cm]{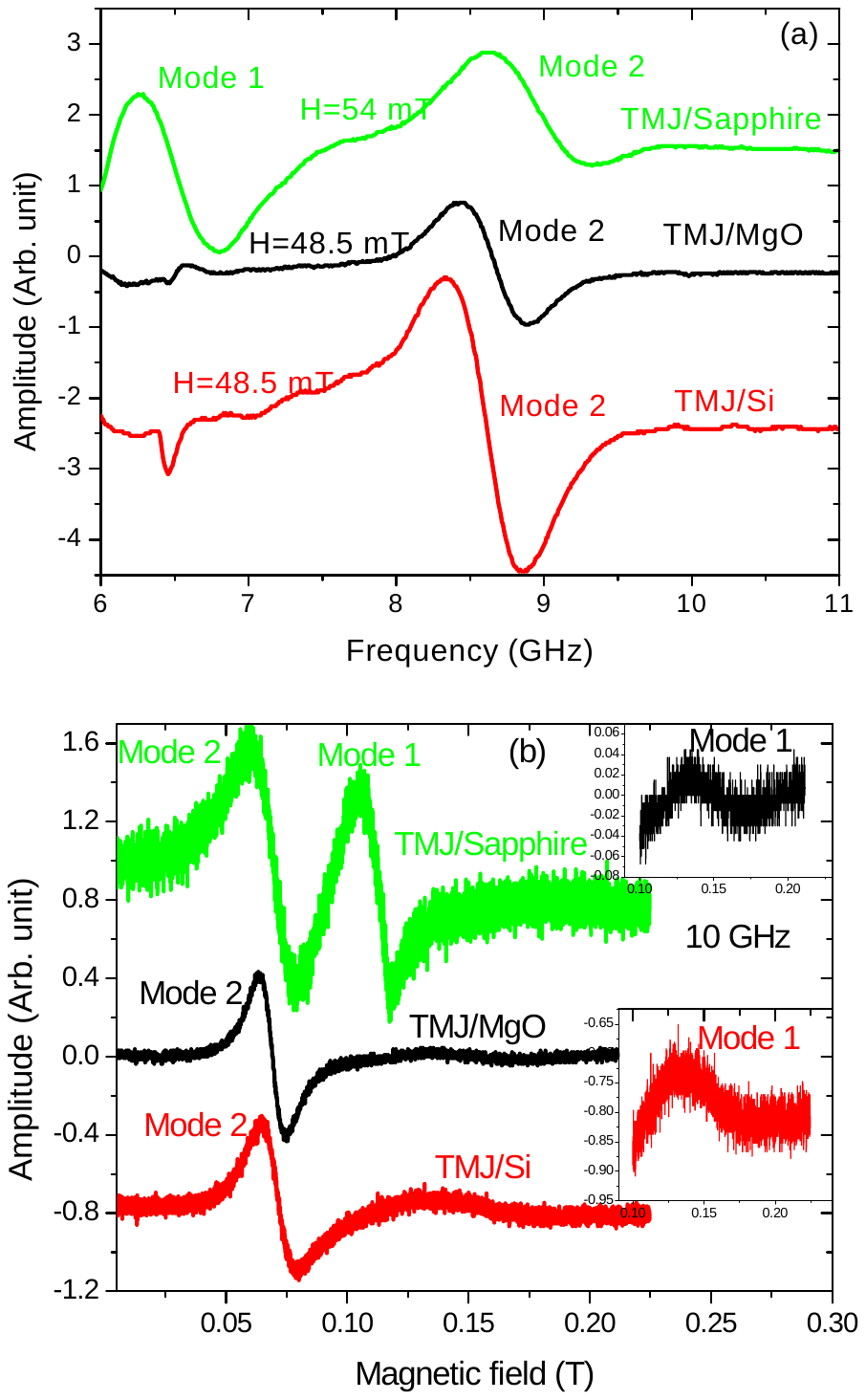}

\caption{(Colour online) (a) sweep frequency and (b) sweep field MS-FMR spectra
of Co$_{2}$MnGe(13nm)/Al$_{2}$O$_{3}$(3nm)/Co(13 nm) TMJs grown
on a-plane sapphire, Si(111) and MgO(100) substrates.}
\end{figure}

\begin{figure}
\includegraphics[bb=20bp 380bp 300bp 595bp,clip,width=8.5cm]{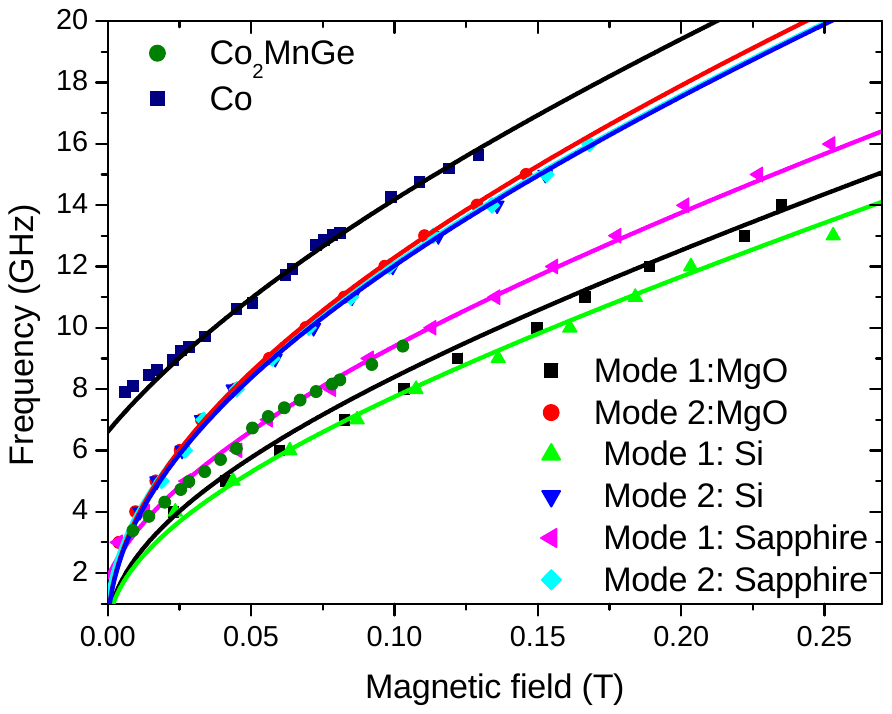}

\caption{(Colour online) Field dependence of the resonance frequency of modes
1 and 2 in Co, Co2MnGe single layers and in Co$_{2}$MnGe(13nm)/Al$_{2}$O$_{3}$(3nm)/Co(13
nm) TMJs grown on a-plane sapphire, on Si(111) and on MgO(100) substrates.
The magnetic field is applied in the film plane. The fits are obtained
considering uncoupled layers using the model presented in the text
with the parameters indicated in Table I.}
\end{figure}

\end{document}